# On Whether People Have the Capacity to Make Observations of Mutually Exclusive Physical Phenomena

Douglas M. Snyder
Los Angeles, California

## Abstract

It has been shown by Einstein, Podolsky, and Rosen that in quantum mechanics either one of two different wave functions predicting specific values for quantities represented by non-commuting Hermitian operators can characterize the same physical existent, without a physical interaction responsible for which wave function occurs. This result means that one can make predictions regarding mutually exclusive features of a physical existent. It is important to ask whether people have the capacity to make observations of mutually exclusive phenomena. Our everyday experience informs us that a human observer is capable of observing only one set of physical circumstances at a time. Evidence from psychology, though, indicates that people may have the capacity to make observations of mutually exclusive physical phenomena, even though this capacity is not generally recognized. Working independently, Sigmund Freud and William James provided some of this evidence. How the nature of the quantum mechanical wave function is associated with the problem posed by Einstein, Podolsky, and Rosen is addressed at the end of the paper.

## Text

In this paper, information has been assembled from a number of sources in physics and psychology in order to explore an issue in quantum mechanical measurement, namely the possibility of measuring mutually exclusive physical phenomena. The resolution of this issue has important implications for psychology as well as physics and, indeed, what their future relationship with one another will be. The problem that occasioned the topic of this paper was initially addressed by Einstein, Podolsky, and Rosen (1935). Their problem concerned the ability in quantum mechanics to make predictions regarding quantities of a physical existent that from a physical standpoint are mutually exclusive. The question is whether human observers have the capacity to make observations that would confirm the mutually exclusive features characterizing the physical existent? The roots of Einstein, Podolsky, and Rosen's problem, as well as the primary concern of this paper, lie in the broad principles of quantum mechanics, in particular as they concern the nature of the wave function associated with a physical entity.

Writings by Sigmund Freud and William James indicate that people may have the capability to make observations on mutually exclusive physical phenomena. These writings are explored. Also, current descriptions of the mental disorders discussed by James are provided as additional evidence to support Freud's and James's conclusions. How the nature of the quantum mechanical wave function is associated with the problem posed by Einstein, Podolsky, and Rosen is addressed at the end of the paper.

QUANTUM MECHANICS

Einstein, Podolsky, and Rosen showed that in quantum mechanics an individual may know quantities of a physical existent that from a physical standpoint are mutually exclusive.[1] An example would be the spin angular momentum components of an electron along orthogonal spatial axes. In one situation, if the spin component of the electron along one of the axes is known precisely (e.g., the $z$ axis), knowledge of the spin component along one of the other axes (e.g., the $y$ axis) is completely uncertain. The spin component along the $y$ axis has a 50-50 chance of having either of two possible values. In another situation, the spin component along the $y$ axis can be known precisely, and the spin component along the $z$ axis is then completely uncertain. The spin component along the $z$ axis in this situation has a 50-50 chance of having either of two possible values.

According to results obtained by Einstein, Podolsky, and Rosen, either of these situations can characterize the electron. Which one does can depend on an event that cannot physically affect the electron. It cannot physically affect the electron because the change in the quantum mechanical wave function occasioned by the event occurs instantaneously throughout space and is therefore not subject to the velocity limitation of the special theory of relativity. The question arises: Do human observers have the capacity to make observations of mutually exclusive physical phenomena, such as those that characterize the electron?

There are a number of points supporting the importance of human observation in quantum mechanical measurement. The first point is that a

---

[1] Bohm (1951) and then Bell (1964) detailed out Einstein, Podolsky, and Rosen's proposal. Experimental evidence (e.g., Aspect, Dalibard, & Roger, 1982) supports this proposal.

# On Whether People

quantum mechanical measurement does not take a final form until a human observer records an observation, due to the ability to consider all of the physical "measurement" interactions as quantum mechanical interactions (Wigner, 1961/1983; Snyder, 1992). A second point is that the occurrence of what are called "negative" observations indicates that a person, in his or her *observational* capacity, is centrally involved in making measurements in quantum mechanics (Bergquist, Hulet, Itano, & Wineland, 1986; Epstein, 1945; Nagourney, Sandberg, & Dehmelt, 1986; Renninger, 1960; Sauter, Neuhauser, Blatt & Toschek, 1986). In negative observations, there is no physical interaction in a measurement of some physical quantity, the physical existent generally changes, and yet a human observer is involved in the measurement process. Thus, the human observer is central to measurement in quantum mechanics.

How can we understand that there then exists the possibility of a person making mutually exclusive observations on a physical existent in quantum mechanics? Our everyday experience informs us that a human observer is capable of observing only one set of physical circumstances at a time.

## EVIDENCE FROM PSYCHOLOGY

Evidence from the discipline of psychology indicates that individuals indeed have the capacity to simultaneously observe mutually exclusive features of a physical existent. Research on the experiential and behavioral adaptation to inversion of incoming light can be combined with experimental scenarios from quantum mechanics involving mutually exclusive physical circumstances to show the aforementioned capacity of human observers. It can be shown for example in analogous circumstances to that usually discussed in the Schrödinger cat gedankenexperimient that essentially Schrödinger's cat can both be alive and dead for different observers and very likely for the same observer (Snyder, 1992, 1993, 1995a, 1995b, 1997).

There is other evidence as well that is relevant to the proposed enhanced observational capacities of humans that has not previously been brought to bear on the issues before us. The evidence in each case is not new, and it stems from the observations of keen observers of the human mind, James and Freud. The quotes that follow are long ones. Both James and Freud stated their positions very well, and the matter before them was subtle as evidenced by Freud's own question on the subject regarding whether the matter of his concern was real or illusory and by James noting how much easier conceptually things would be if mutually exclusive selves and/or consciousnesses were not supported by empirical data.

The psychological tendency in people to maintain a sense of wholeness and integrity is a strong one.  It has been considered, for example, a hallmark of mental health.  Yet the data indicate that the mind has a larger capability for maintaining diverse, or mutually exclusive, viewpoints at the same time.  It is this feature of the mind that both James and Freud were concerned with in the work to be discussed.  They saw evidence of it in those diagnosed as having a mental illness, and they extended the results of their investigations with these individuals to those who are normal.

It should be emphasized that these mutually exclusive viewpoints may exist simultaneously.  Yet, they maintain some connection to each other.  That is they affect one another, and it may be said that each would not exist without the other.  But this is very different than stating that these viewpoints are really simply modes of expression of a unified consciousness.  Instead they are entities existing in a common world and thus maintaining certain relations with one another, but nonetheless existing as distinct and separate entities.

Near the end of his life, Freud (1940a/1964) wrote a description of key elements of psychoanalysis entitled *An Outline of Psychoanalysis*.  In a section of this work entitled "The External World," he took up a topic he had begun in a paper that he had written about a few years before but which he never completed.  In this earlier paper, "Splitting of the Ego in the Process of Defence," Freud (1940b/1964) began by wondering whether the topic he was about to discuss was really of significance or whether he had explored it in depth before. This statement is quite curious.  In this paper and in *An Outline of Psychoanalysis*, Freud attempted to grapple in a new way with the issue of there being two distinct psychological elements working within an individual, oftentimes working at odds with one another.  This, of course, seems to be a basic premise of psychoanalysis and generally falls under the rubric of intrapsychic dynamics.  Indeed, the direction of an individual's personality after childhood is in Freud's psychoanalysis determined by the relative balance of energy at the disposal of the ego on the one hand and other psychological structures such as the id and super-ego as well as the demands put on the ego by the external world.

Freud's curiosity though was drawn anew to this incongruity that a single individual could simultaneously have these conflicting psychological elements.  He approached the topic of mutually exclusive situations involving the mind initially from areas where this mutual exclusivity is readily apparent and then proceeded to work toward the less extreme types of mental disorder and finally to normal psychological functioning.  Freud's discussion is compelling and points toward the need to systematically explore the potential in individuals to simultaneously manifest mutually exclusive modes of psychological functioning.  What generally passes as "normality" in our own experience masks these distinct modes in some sort of integration or unification.  But this "integration" is not actually a fusing of the mutually exclusive modes.  It is more of an enveloping of the modes as the ego exerts as much effort as it can to enwrap them in a skin that makes them appear integrated, consistent, and understandable.  Thus the act of eating for example can express both destructive and constructive features of the mind and be seen as a single act, though these mutually exclusive features cannot reduced to a single feature of the mind.  Psychopathological conditions, though, present situations where a single act does not allow for fusing, or "integration" of, these different psychological features and pointedly shows that the veneer of unification in everyday experience masks the fundamental and simultanaeous presence of mutually exclusive psychological features.

In *The Principles of Psychology*, James (1899) suggested the same point concerning "integration" in discussing certain behaviors of hysterics that today would be found in individuals diagnosed with Conversion Disorder and/or Dissociative Identity Disorder.

The simultaneously existing, mutually exclusive modes of psychological functioning  have their own perceptual systems.  This being the case, Freud's and James's work is significant to the problem explored in this paper.  Though their work was concerned largely with mental disorder, there is a current running through it indicating that the mutually exclusive modes of psychological functioning are general factors, not limited to mental disorder.  The denial that may characterize one perceptual mode in some forms of mental illness turns into different kinds and levels of attention, or different forms of adaptation, concerning features of the physical world for the mutually exclusive modes of functioning in the "normal" mind.  The mutually exclusive perceptual phenomena that the mind can support can be tied to the mutually exclusive features of a physical existent that can occur in quantum mechanics.

First, a portion of "The External World" from Freud's *An Outline of Psychoanalysis* is presented.  His unfinished paper, "Splitting of the Ego in

# On Whether People

Defence" is then presented. From *The Principles of Psychology*, part of the chapter entitled "The Relation of Minds to Other Things" is presented.

### AN OUTLINE OF PSYCHOANALYSIS: THE EXTERNAL WORLD

According to Freud, the ego is the psychological structure that has the executive duties for mental functioning. It manages the conflicting demands exerted by other psychological structures and those imposed by the external world. In its executive role, the ego is responsible for rational thought as well as consciousness.

> We have repeatedly had to insist on the fact that the ego owes its origin as well as the most important of its acquired characteristics to its relation to the real external world. We are thus prepared to assume that the ego's pathological states, in which it most approximates once again to the id, are founded on cessation or slackening of that relation to the external world. This tallies very well with what we learn from clinical experience–namely, that the precipitating cause of the outbreak of psychosis is either that reality has become intolerably painful or that the instincts have become extraordinarily intensified–both of which, in view of the rival claims made on the ego by the id and the external world, must lead to the same result. (p. 201)

This is a straightforward view of psychoses in which the ego is simply too weak to stand up to and manage the demands of the id or the external world and the individual simply withdraws from rational interaction with the physical world. The ego, which is the psychological structure that interfaces with the external world, is overwhelmed. The pleasure principle dominates psychological functioning according to which the individual seeks immediate gratification without distinguishing whether the source of the gratification is real or illusory.

> The problem of psychoses would be simple and perspicuous if the ego's detachment from reality could be carried through completely. (p. 201)

But Freud says the situation is not so simple. Rather we get a situation like two executive structures, two egos each functioning independently of the other. Freud discussed this situation in the case of psychoses.

# On Whether People

But that seems to happen only rarely or perhaps never. *Even in a state so far removed from the reality of the external world as one of hallucinatory confusion, one learns from patients after their recovery that at the time in some corner of their mind (as they put it) there was a normal person hidden, who, like a detached spectator, watched the hubbub of illness go past him* [italics added]. I do not know if we may assume that this is so in general, but I can report the same of other psychoses with a less tempestuous course. I call to mind a case of chronic paranoia in which after each attack of jealousy a dream conveyed to the analyst a correct picture of the precipitating cause, free from any delusion. An interesting contrast was thus brought to light: while we are accustomed to discover from the dreams of neurotics jealousies which are alien to their waking lives, in this psychotic case the delusion which dominated the patient in the day-time was corrected by his dream. *We may probably take it as being generally true that what occurs in all these cases is a psychical split. Two psychical attitudes have been formed instead of a single one–one, the normal one which takes account of reality, and another which under the influence of the instincts detaches the ego from reality. The two exist alongside of each other* [italics added except for "split"]. The issue depends on their relative strength. If the second is or becomes the stronger, the necessary precondition for a psychosis is present. If the relation is reversed, then there is an apparent cure of the delusional disorder. Actually it has only retreated into the unconscious just as numerous observations lead us to believe that the delusion existed ready-made for a long time before its manifest irruption. (. p. 201-202)

Freud then extended his discussion of the splitting of the ego to other psychopathological conditions. He began by discussing fetishes.

*The view which postulates that in all psychoses there is a splitting of the ego could not call for so much notice if it did not turn out to apply to other states more like the neuroses and, finally, to the neuroses themselves* [italics added except for "splitting of the ego"]. I first became convinced of this in cases of *fetishism*. This abnormality, which may be counted as one of

> the perversions, is, as is well known, based on the patient (who is almost always male) not recognizing the fact that females have no penis–a fact which is extremely undesirable to him since it is a proof of the possibility of his being castrated himself.  He therefore disavows his own sense-perception which showed him that the female genitals lack a penis and holds fast to the contrary conviction.  The disavowed perception does not, however, remain entirely without influence for, in spite of everything, he has not the courage to assert that he actually saw a penis.  He takes hold of something else instead–a part of the body or some other object–and assigns it the role of the penis which he cannot do without.  It is usually something that he in fact saw at the moment at which he saw the female genitals, or it is something that can suitably serve as a symbolic substitute for the penis.  Now it would be incorrect to describe this process when a fetish is constructed as a splitting of the ego; it is a compromise formed with the help of displacement, such as we have been familiar with in dreams. (pp. 202-203)

It is a compromise where the underlying concern to the individual is unconscious and yet there is some allowance for his concern and the perception accompanying it, namely the existence of a penis in a woman.

> But our observations show us still more.  The creation of the fetish was due to an intention to destroy the evidence for the possibility of castration, so that fear of castration could be avoided.  If females, like other living creatures, possess a penis, there is no need to tremble for the continued possession of one's own penis. (p. 203)

Here we have the origin of the development of two executive functions, the splitting of the ego.  Also, note that there is a consistent and thorough basis for the development of an alternative executive function.  The individual's intention is to get rid of the possibility of castration, an attempt that cannot wholly succeed because according to Freud, this possibility is a basic element of psychosexual development.  Thus we arrive at the beginning of two mutually exclusive attitudes toward the same phenomenon.  Note though that this split ego nonetheless relies on an some acknowledgement by at least one part of the ego (in this case, the part of the ego that disavows the individual's perception) of the other executive function that together with the former constitute the split ego.

Then Freud showed that we indeed have two executive functions that appear to be functioning independently of each other. The "normal" executive function develops independently of the "abnormal" one.

> *Now we come across fetishists who have developed the same fear of castration as non-fetishists and react in the same way to it. Their behaviour is therefore simultaneously expressing two contrary premisses* [italics added]. On the one hand they are disavowing the fact of their perception–the fact that they saw no penis in the female genitals; and on the other hand they are recognizing the fact that females have no penis and are drawing the correct conclusions from it. (p. 203)

Note that both attitudes involve perception and thus the involvement of the ego in both is essential. The ego is split, maintaining contrary attitudes in response to the fear generated by sexual impulses seeking uninhibited expression.

> *The two attitudes persist side by side throughout their lives without influencing each other. Here is what may rightly be called a splitting of the ego* [italics added]. This circumstance also enables us to understand how it is that fetishism is so often only partially developed. It does not govern the choice of object exclusively but leaves room for a greater or lesser amount of normal sexual behaviour; sometimes, indeed, it retires into playing a modest part or is limited to a mere hint. In festishists, therefore, the detachment of the ego from the reality of the external world has never succeeded completely. (p. 203)

Freud then proceeded one step further by showing how the splitting of the ego is not limited to psychoses and fetishes. He showed how in the general process of psychological development, an individual may disavow aspects of their perceptions that ameliorate some demand being made on a child. Note that these demands from the external world assume importance in large measure because of instinctual demands that are not acceptable in the external world. Freud wrote that this disavowal is at the heart of the development of the two independently functioning executive functions.

# On Whether People

> It must not be thought that fetishism presents an exceptional case as regards a splitting of the ego; it is merely a particularly favourable subject for studying the question. Let us return to our thesis that the childish ego, under the domination of the real world, gets rid of undesirable instinctual demands by what are called repressions. We will now supplement this by further asserting that, during the same period of life, the ego often enough finds itself in the position of fending off some demand from the external world which it feels distressing and that this is effected by means of a disavowal of the perceptions which bring to knowledge this demand from reality. Disavowals of this kind occur very often and not only with fetishists; and whenever we are in a position to study them they turn out to be half measures, incomplete attempts at detachment from reality. The disavowal is always supplemented by an acknowledgement; two contrary and independent attitudes always arise and result in the situation of there being a splitting of the ego. Once more the issue depends on which of the two can seize hold of the greater intensity. (pp. 203-204)

Freud then noted again that the process of the splitting of the ego, is not uncommon to psychological development. Freud then noted the existence of distinct and opposing attitudes that are represented in behaviors of the neurotic. These are found in neurotic symptoms.

> *The facts of this splitting of the ego, which we have just described, are neither so new nor so strange as they may at first appear. It is indeed a universal characteristic of neuroses that there are present in the subject's mental life, as regards some particular behaviour, two different attitudes, contrary to each other and independent of each other. In the case of neuroses, however, one of these attitudes belongs to the ego and the contrary one, which is repressed, belongs to the id* [italics added]. (p. 204)

Freud then noted that neurosis and, I believe, fetishism are different topographically or structurally regarding the splitting of the ego, not in terms of process. He noted that they both involve compromise between two distinct and opposing attitudes without fully distinguishing what the essential difference is. Freud implied that as far as our general awareness is concerned, individuals function with a unified sense of our experience and that the simultaneous existence of the distinct and opposing attitudes is not what we generally feel. If we become aware of such attitudes, they generally are in a sequence, one at a time, not all at the same time.

> The difference between this case and the other [discussed in the previous paragraph] is essentially a topographical or structural one, and it is not always easy to decide in an individual instance with which of the two possibilities one is dealing. They have, however, the following important characteristic in common. Whatever the ego does in its efforts of defence, whether it seeks to disavow a portion of the real external world or whether it seeks to reject an instinctual demand from the internal world, its success is never complete and unqualified. The outcome always lies in two contrary attitudes, of which the defeated, weaker one, no less than the other, leads to psychical complications. In conclusion, it is only necessary to point out how little of all these processes becomes known to us through our conscious perception [where we act with a unified sense of experience]. (p. 204)[2]

### SPLITTING OF THE EGO IN THE PROCESS OF DEFENSE

According to Strachey (1964), this paper was written shortly before *An Outline of Psychoanalysis*.[3] Freud (1940b/1964) began this work by noting that he was unsure whether the conflicting attitudes in neurosis, and in normal behavior as well, are fundamentally different than that found for a splitting of the ego in psychoses and fetishes.

> I find myself for a moment in the interesting position of not knowing whether what I have to say should be regarded as something long familiar and obvious or as something entirely new and puzzling. But I am inclined to think the latter.

---

[2] *[discussed in the previous paragraph]* is from the original text.

[3] "Splitting of the Ego in the Process of Defence" is presented after the text from *An Outline of Psychoanalysis* because in the latter work Freud extends his notion of the splitting of the ego to neurosis and and by implication to general psychological functioning. Strachey (1964) wrote that in *An Outline of Psychoanalysis*, Freud "extends the application of the idea of a splitting of the ego beyond the cases of fetishism and of the psychoses to neuroses in general. Thus the topic links up with the wider question of the 'alteration of the ego' which is invariably brought about by the processes of defence" (p. 274).

> I have at last been struck by the fact that the ego of a person whom we know as a patient in analysis must, dozens of years earlier, when it was young, have behaved in a remarkable manner in certain particular situations of pressure. We can assign in general and somewhat vague terms the conditions under which this comes about, by saying that it occurs under the influence of a psychical trauma. I prefer to select a single sharply defined special case, though it certainly does not cover all the possible modes of causation. (p. 275)

Freud began by talking about the general process of development that may lead to psychopathology. The distinguishing characteristic of psychopathology for Freud is that the instinctual demand is stronger than the capability of the ego to manage it in the face of reality.

> Let us suppose, then, that a child's ego is under the sway of a powerful instinctual demand which it is accustomed to satisfy and that it is suddenly frightened by an experience which teaches it that the continuance of this satisfaction will result in an almost intolerable real danger. It must now decide either to recognize the real danger, give way to it and renounce the instinctual satisfaction, or to disavow reality and make itself believe that there is no reason for fear, so that it may be able to retain the satisfaction. Thus there is a conflict between the demand by the instinct and the prohibition by reality. (p. 275)

Here is the prototypical developmental situation confronting a child where the desire for instinctual satisfaction must be managed because of perceived negative consequences from the environment that will result from the continuation of behavior directed toward this satisfaction. The path to psychosis lies in a strong disavowal of reality. The path to psychological maturity and to neurosis lies in reducing instinctual satisfaction.[4]

---

[4] These alternatives toward satisfying the demands of the environment or of instinct actually blend with one another in the individual. There may well be a simple denial of some event while intact reality testing is maintained. There is very often instinctual satisfaction where the primary reaction of the individual is to control it and even to minimize it, particularly its primitive expression.

# On Whether People

But Freud was headed somewhere else, somewhere that can only be found by a more subtle consideration of psychodynamics. Freud presents the situation where two executive agencies take different approaches toward handling the drive to instinctual expression in the face of limitations imposed by the environment. One of these agencies appears in some ways as the ego that is too weak to stand up to the instinctual demands. This agency develops a symptom, the fetish, that allows for some disguised primitive sexual expression. The other agency appears like the normal ego engaging in normal mature sexual expression. But this agency is anything but normal, relying on the other one to provide the "cover," a way of dealing with the fear generated by the desired instinctual expression so that this agency can go on its merry way without being aware of this fear. This situation Freud refers to as a splitting of the ego. The executive agencies are intertwined, but yet for all intents and purposes are also independent as each embodies what generally is considered either a normal, healthy ego or a neurotic one.

> But in fact the child takes neither course, or rather he takes both simultaneously, which comes to the same thing. He replies to the conflict with two contrary reactions, both of which are valid and effective. On the one hand, with the help of certain mechanisms he rejects reality and refuses to accept any prohibition; on the other hand, in the same breath he recognizes the danger of reality, takes over the fear of that danger as a pathological symptom and tries subsequently to divest himself of the fear. It must be confessed that this is a very ingenious solution of the difficulty. Both of the parties to the dispute obtain their share: the instinct is allowed to retain its satisfaction and proper respect is shown to reality. But everything has to be paid for in one way or another, and this success is achieved at the price of a rift in the ego which never heals but which increases as time goes on. The two contrary reactions to the conflict persist as the centre-point of a splitting of the ego. The whole process seems so strange to us because we take for granted the synthetic nature of the processes of the ego. But we are clearly at fault in this. The synthetic function of the ego, though it is of such extraordinary importance, is subject to particular conditions and is liable to a whole number of disturbances. (pp. 275-276)

# On Whether People

Freud then introduced the specific features of the case history he presented that illustrates how the general principles discussed in the previous quoted paragraph may be manifested.

> It will assist if I introduce an individual case history into this schematic disquisition. A little boy, while he was between three and four years of age, had become acquainted with the female genitals through being seduced by an older girl. After these relations had been broken off, he carried on the sexual stimulation set going in this way by zealously practicing manual masturbation; but he was soon caught at it by his energetic nurse and was threatened with castration, the carrying out of which was, as usual, ascribed to his father. There were thus present in this case conditions calculated to produce a tremendous effect of fright. A threat of castration by itself need not produce a great impression. A child will refuse to believe in it, for he cannot easily imagine the possibility of losing such a highly prized part of his body. His [earlier] sight of the female genitals might have convinced our child of that possibility. But he drew no such conclusion from it, since his disinclination to doing so was too great and there was no motive present which could compel him to. On the contrary, whatever uneasiness he may have felt was calmed by the reflection that what was missing would yet make its appearance: she would grow one (a penis) later. Anyone who has observed enough small boys will be able to recollect having come across some such remark at the sight of a baby sister's genitals. But it is different if both factors are present together. In that case the threat revives the memory of the perception which had hitherto been regarded as harmless and finds in that memory a dreaded confirmation. The little boy now thinks he understands why the girl's genitals showed no sign of a penis and no longer ventures to doubt that his own genitals may meet with the same fate. Thenceforward he cannot help believing in the reality of the danger of castration. (pp. 276-277)

> The usual result of the fright of castration, the result that passes as the normal one, is that, either immediately or after some considerable struggle, the boy gives way to the threat and obeys the prohibition either wholly or at least in part (that is, by

> no longer touching his genitals with his hand). In other words, he gives up, in whole or in part, the satisfaction of the instinct. We are prepared to hear, however, that our present patient found another way out. He created a substitute for the penis which he missed in females–that is to say, a fetish. In so doing, it is true that he had disavowed reality, but he had saved his own penis. So long as he was not obliged to acknowledge that females have lost their penis, there was no need for him to believe the threat that had been made against him: he need have no fears for his own penis, so he could proceed with his masturbation undisturbed. (p. 277)

Freud noted that the fetish involved a loosening of the tie to reality, but unlike the psychoses, involved a displacement of value only.

> This behaviour on the part of our patient strikes us forcibly as being a turning away from reality–a procedure which we should prefer to reserve for psychoses. And it is in fact not very different. Yet we will suspend our judgement, for upon closer inspection we shall discover a not unimportant distinction. The boy did not simply contradict his perceptions and hallucinate a penis where there was none to be seen; he effected no more than a displacement of value–he transferred the importance of the penis to another part of the body, a procedure in which he was assisted by the mechanism of regression (in a manner which need not here be explained). This displacement, it is true, related only to the female body; as regards his own penis nothing was changed. (p. 277)

Freud wrote the final part of this unfinished piece. He discussed the widely differing behaviors reflective of the functioning of two independent psychological agencies.

> This way of dealing with reality, which almost deserves to be described as artful, was decisive as regards the boy's practical behaviour. He continued with his masturbation as though it implied no danger to his penis; but at the same time, in complete contradiction to his apparent boldness or indifference, he developed a symptom which showed that he nevertheless did recognize the danger. He had been threatened with being castrated by his father, and immediately afterwards, simultaneously with the creation of his fetish, he developed an intense fear of his father punishing him, which it required the whole force of his masculinity to master and overcompensate. This fear of his father, too, was silent on the subject of castration: by the help of regression to an oral phase, it assumed the form of a fear of being eaten by his father. At this point it is impossible to forget a primitive fragment of Greek mythology which tells how Kronos, the old Father God, swallowed his children and sought to swallow his youngest son Zeus like the rest, and how Zeus was saved by the craft of his mother and later on castrated his father. But we must return to our case history and add that the boy produced yet another symptom, though it was a slight one, which he has retained to this day. This was an anxious susceptibility against either of his little toes being touched, as though, in all the to and fro between disavowal and acknowledgement, it was nevertheless castration that found the clearer expression.... (pp. 277-278)

Now James's work will be explored concerning the possibility of the mind simultaneously manifesting what James referred to as mutually exclusive consciousnesses.

### THE PRINCIPLES OF PSYCHOLOGY

Immediately before beginning his section of "Unconsciousness in Hysterics," James (1899) wrote:

*a lot of curious observations made on hysterical and hypnotic subjects, which prove the existence of a highly developed consciousness in places where it has hitherto not been suspected at all. These observations throw such a novel light upon human nature that I must give them is some detail. That at least four different and in a certain sense rival observers should agree in the same conclusion justifies us in the accepting the conclusion as true* [italics added]. (p. 202)[5]

---

[5] James listed more than four individuals in the material presented here who could meet this criterion. Among them are Pierre Janet, Jules Janet, Binet, Gurney, Beaunis, Bernheim, and Pitres.

# On Whether People

*'Unconsciousness' in Hysterics*

James began by describing a common symptom of severe cases of what was known as hysteria, namely some alteration of the body's sensibility that did not correspond to a meaningful organic disease pattern.

> One of the most constant symptoms in persons suffering from hysteric disease in its extreme forms consists in alterations of the natural sensibility of various parts and organs of the body. Usually the alteration is in the direction of defect, or anæsthesia. One or both eyes are blind, or color-blind, or there is hemianopsia (blindness to one half the field of view), or the field is contracted. Hearing, taste, smell may similarly disappear, in part or in totality. Still more striking are the cutaneous anæsthesias. The old witch-finders looking for the 'devil's seals' learned well the existence of those insensible patches on the skin of their victims, to which the minute physical examinations of recent medicine have but recently attracted attention again. They may be scattered anywhere, but are very apt to affect one side of the body. Not infrequently they affect an entire lateral half, from head to foot; and the insensible skin of, say, the left side will then be found separated from the naturally sensitive skin of the right by a perfectly sharp line of demarcation down the middle of the front and back. Sometimes, most remarkable of all, the entire skin, hands, feet, face, everything, and the mucous membranes, muscles and joints so far as they can be explored, become *completely* insensible without the other vital functions becoming gravely disturbed. (pp. 202-203)

James explained the ways known at the time that could make hysterical anesthesia disappear, in particular by noting the use of a hypnotic trance.

> These hysterical anæsthesias can be made to disappear more or less completely by various odd processes. It has been recently found that magnets, plates of metal, or the electrodes of a battery, placed against the skin, have this peculiar power. And when one side is relieved in this way, the anæsthesia is often found to have transferred itself to the opposite side, which until then was well. Whether these strange effects of magnets and

> metals be due to their direct physiological action, or to a prior effect on the patient's mind ('expectant attention' or 'suggestion') is still a mooted question. A still better awakener of sensibility is the hypnotic trance, into which many of these patients can be very easily placed, and in which their lost sensibility not infrequently becomes entirely restored. Such returns of sensibility succeed the times of insensibility and alternate with them. (p. 203)

Then James went one step further and noted that hysterical anesthesia and its absence, alternating in some kind of cyclical pattern, need not be the only scheme of presentation of these different states of consciousness. Also, these different states may be simultaneous, giving rise to James's contention that there may exist different forms of consciousnesses simultaneously which are independent of one another.

> But Messrs. Pierre Janet and A. Binet have shown that during the times of anæsthesia, and coexisting with it, *sensibility to the anæesthetic parts is also there, in the form of a secondary consciousness* entirely cut off from the primary or normal one, but susceptible of being *tapped* and made to testify to its existence in various odd ways. (p. 203)

James noted that hysterics may have a very limited scope of attention and that these individuals are open to a "method of distraction" that allows both consciousnesses to express themselves simultaneously.

> Chief amongst these is what M. Janet calls 'the method of *distraction*.' These hysterics are apt to possess a very narrow field of attention, and to be unable to think of more than one thing at a time. When talking with any person they forget everything else. "When Lucie talked directly with anyone," says M. Janet, "she ceased to be able to hear any other person. You may stand behind her, call her by name, shout abuse into her ears, without making her turn round; or place yourself before her, show her objects, touch her, etc., without attracting her notice. When finally she becomes aware of you, she thinks you have just come into the room again, and greets you accordingly. This singular forgetfulness makes her liable to tell all her secrets aloud, unrestrained by the presence of unsuitable auditors." (p. 203)

James noted how Janet used this "method of distraction" to show the simultaneous existence of different, independent consciousnesses. Note that here the same sense, hearing, is being used by both consciousnesses simultaneously.

> Now M. Janet found in several subjects like this that if he came up behind them whilst they were plunged in conversation with a third party, and addressed them in a whisper, telling them to raise their hand or perform other simple acts, they would obey the order given, although their *talking* intelligence was quite unconscious of receiving it. Leading them from one thing to another, he made them reply by signs to his whispered questions, and finally made them answer in writing, if a pencil were placed in their hand. The primary consciousness meanwhile went on with the conversation, entirely unaware of these performances on the hand's part. The consciousness which presided over these latter appeared in its turn to be quite as little disturbed by the upper consciousness's concerns. This *proof by 'automatic' writing*, of a secondary consciousness's existence, is the most cogent and striking one; but a crowd of other facts prove the same thing. If I run through them rapidly, the reader will probably be convinced. (p. 204)

James proceeded to discuss ways in which there could be a psychological split in perception that underlaid the simultaneous existence of mutually exclusive situations. Note the similarity to Freud's example of a splitting of the ego where one executive agency appears normal but is in fact not because of its co-existence, and relationship, with a clearly disordered executive agency. Also, note that the sense of touch is being used by both consciousnesses simultaneously. One claimed that it cannot feel with one hand. The other relied on the sense of touch in that same hand to adjust to the object in it.

> *The apparently anæsthetic hand* of these subjects, for one thing, *will often adapt itself discriminatingly* to whatever object may be put into it. With a pencil it will make writing movements; into a pair of scissors it will put its fingers and will open and shut them, etc., etc. The primary consciousness, so to call it, is meanwhile unable to say whether or no [sic] *anything* is in the hand, if the latter be hidden from sight. "I put a pair of eyeglasses into Léonie's anæsthetic hand, this hand opens it and raises it towards the nose, but half way thither it enters the field of vision of Léonie, who sees it and stops stupefied: 'Why,' says she, 'I have an eyeglass in my left hand!'" M. Binet found a very curious sort of connection between the apparently anæsthetic skin and the mind in some Salpétrière-subjects. Things placed in the hand were not felt, but *thought* of (apparently in visual terms) and in no wise referred by the subject to their starting point in the hand's sensation. A key, a knife, placed in the hand occasioned *ideas* of a key or a knife, but the hand felt nothing. Similarly the subject *thought* of the number 3, 6, etc., if the hand or finger was bent three or six times by the operator, or if he stroked it three, six, etc., times. (p. 204)

James also pointed out that one consciousness may have certain ideas due to the other consciousness having had a certain experience, with the former not knowing about the latter. James provided more examples of the phenomenon of one consciousness being affected by an experience of the other without the former consciousness knowing of the latter's experience.

In certain individuals there was found a still odder phenomenon, which reminds one of that curious idiosyncrasy of 'colored hearing' of which a few cases have been lately described with great care by foreign writers. These individuals, namely, *saw* the impression received by the hand, but could not feel it; and the thing seen appeared by no means associated with the hand, but more like an independent vision, which usually interested and surprised the patient. Her hand being hidden by a screen, she was ordered to look at another screen and to tell of any visual image which might project itself thereon. Numbers would then come, corresponding to the number of times the insensible member was raised, touched, etc. Colored lines and figures would come, corresponding to similar ones traced on the palm; the hand itself or its fingers would come when manipulated; and finally objects placed in it would come; but on the hand itself nothing would ever be felt. Of course simulation would not be hard here; but M. Binet disbelieves this (usually very shallow) explanation to be a probable one in cases in question. (p. 205)

In a footnote on this last quote, James shows clearly that he maintained that more than one self may exist for a person.

*This whole phenomenon shows how an idea which remains itself below the threshold of a certain conscious self* [italics added] may occasion associative effects therein. The skin-sensations unfelt by the patients primary consciousness awaken nevertheless their usual visual associates therein. (p. 205)

James continued by explicitly pointing out differences in perception of the same stimuli among the various consciousnesses that exist at the same time for the same sensory modality as evidenced by explicit responses from the subject.

The usual way in which doctors measure the delicacy of our touch is by the compass-points. Two points are normally felt as one whenever they are too close together for discrimination; but what is 'too close' on one part of the skin may seem very far apart on another. In the middle of the back or on the thigh, less than 3 inches may be too close; on the finger-tip a tenth of an inch is far enough apart. Now, as tested in this way, with the appeal made to the primary consciousness, which talks through the mouth and seems to hold the field alone, a certain person's skin may be entirely anæsthetic and not feel the compass-points at all; and yet this same skin will prove to have a perfectly normal sensibility if the appeal be made to that other secondary or sub-consciousness, which expresses itself automatically by writing or by movements of the hand. M. Binet, M. Pierre Janet, and M. Jules Janet have all found this. The subject, whenever touched, would signify 'one point' or 'two points,' as accurately as if she were a normal person. She would signify it only by these movements; and of the movements themselves her primary self would be as unconscious as of the facts they signified, for what the submerged consciousness makes the hand do automatically is unknown to the consciousness which uses the mouth. (pp. 205-206)

# On Whether People

James discussed similar phenomena to that presented for discriminative touch in the case of visual perception.

> Messrs. Bernheim and Pitres have also proved, by observations, too complicated to be given in this spot, that the hysterical blindness is no real blindness at all. The eye of an hysteric which is totally blind when the other or seeing eye is shut, will do its share of vision perfectly well when both eyes are open together. But even where both eyes are semi-blind from hysterical disease, the method of automatic writing proves that their perceptions exist, only cut off from communication with the upper consciousness. M. Binet has found the hand of his patients unconsciously writing down words which their eyes were vainly endeavoring to 'see,' i.e., to bring to the upper consciousness. Their submerged consciousness was of course seeing them, or the hand could not have written as it did. Colors are similarly perceived by the sub-conscious self, which the hysterically color-blind eyes cannot bring to the normal consciousness. Pricks, burns, and pinches on the anæsthetic skin, all unnoticed by the upper self, are recollected to have been suffered, and complained of, as soon as the under self gets a chance to express itself by the passage of the subject into hypnotic trance. (p. 206)

James then summarized the results and stated his conclusion that an individual may show mutually exclusive consciousnesses that exist simultaneously. Importantly, he noted that these consciousnesses were not aware of the same object at the same time. He called this characteristic "complementary." He exemplified this with the case of the post-hypnotic trance behavior of Lucie.

> *It must be admitted, therefore, that in certain persons, at least, the total possible consciousness may be split into parts which coexist but mutually ignore each other, and share the objects of knowledge between them. More remarkable still, they are complementary. Give an object to one of the consciousnesses, and by that fact you remove it from the other or others* [italics added]. Barring a certain common fund of information, like the command of language, etc., what the upper self knows the under self is ignorant of, and *vice versa*. (p. 206)[6]

James then provided an example of the conclusion that he just drew that led to his discussion of a second self, or executive agency, besides the one that is shown publicly.

> M. Janet has proved this beautifully in his subject Lucie. The following experiment will serve as the type of the rest: In her trance he covered her lap with cards, each bearing a number. He then told her that on waking she should not see any card whose number was a multiple of three. This is the ordinary so-called 'post-hypnotic suggestion,' now well known, and for which Lucie was a well-adapted subject. Accordingly, when she was awakened and asked about the papers on her lap, she counted and said she saw those only whose number was not a multiple of 3. To the 12, 18, 9, etc., she was blind. But the hand, when the sub-conscious self was interrogated by the usual method of engrossing the upper self in another conversation, wrote that the only cards in Lucie's lap were those numbered 12, 18, 9, etc., and on being asked to pick up all the cards which were there, picked up these and let the others lie. Similarly when the sight of certain things was suggested to the sub-conscious Lucie, the normal Lucie suddenly became partially or totally blind. "What is the matter? I can't see!" the normal personage suddenly cried out in the midst of her conversation, when M. Janet whispered to the secondary personage to make use of her eyes. The anæsthesia's, paralyses, contractions and other irregularities from which hysterics suffer seem then to be due to the fact that their secondary personage has enriched itself by robbing the primary one of a function which the latter ought to have retained. (pp. 206-207)

James then indicated how Jules Janet attempted to resolve the symptoms of a patient that were under the control of the less accessible executive structure.

---

[6] "*In certain persons, at least, the total possible consciousness may be split into parts which coexist but mutually ignore each other*" and *"complementary"* are italicized in the original text.

# On Whether People

> The curative indication is evident: get at the secondary personage, by hypnotization or in whatever other way, and make her *give up* the eye, the skin, the arm, or whatever the affected part may be. The normal self thereupon regains possession, sees, feels, or is able to move again. In this way M. Jules Janet easily cured the well known subject of the Salpêtrière, Witt., of all sorts of afflictions which, until he discovered the secret of her deeper trance, it had been difficult to subdue. "Cessez cette mauvaise plaisanterie," he said to the secondary self–and the latter obeyed. The way in which the various personages share the stock of possible sensations between them seems to be amusingly illustrated in this young woman. When awake, her skin is insensible everywhere except on a zone about the arm where she habitually wears a gold bracelet. This zone has feeling; but in the deepest trance, when all the rest of her body feels, this particular zone becomes absolutely anæsthetic. (p. 207)

James provided an example of the incongruent sets of coordinated behaviors that an individual may display, providing support for the existence of mutually exclusive consciousnesses that simultaneously exist.

> Sometimes the mutual ignorance of the selves leads to incidents which are strange enough. The acts and movements performed by the sub-conscious self are withdrawn from the conscious one, and the subject will do all sorts of incongruous things of which he remains quite unaware. "I order Lucie [by the method of *distraction*] to make a *pied de nez*, and her hands go forthwith to the end of her nose. Asked what she is doing, she replies that she is doing nothing, and continues for a long time talking, with no apparent suspicion that her fingers are moving in front of her nose. I make her walk about the room; she continues to speak and believes herself sitting down." (p. 208)

James provided other examples, examples that show the degree to which an individual may go to maintain the incongruent sets of behaviors and his own witnessing such behaviors.

# On Whether People

> M. Janet observed similar acts in a man in alcoholic delirium. Whilst the doctor was questioning him, M. J. made him by whispered suggestion walk, sit, kneel, and even lie down on his face on the floor, he all the while believing himself to be standing beside his bed. Such *bizarreries* sound incredible, until one has seen their like. Long ago, without understanding it, I myself saw a small example of the way in which a person's knowledge may be shared by the two selves. A young woman who had been writing automatically was sitting with a pencil in her hand, trying to recall at my request the name of a gentleman whom she had once seen. She could only recollect the first syllable. Her hand meanwhile, without her knowledge, wrote down the last two syllables. In a perfectly healthy young man who can write with the planchette, I lately found the hand to be entirely anæsthetic during the writing act; I could prick it severely without the Subject knowing the fact. The *writing on the planchette*, however, accused me in strong terms of hurting the hand. Pricks on the *other* (non-writing) hand, meanwhile, which, awakened strong protest from the young man's vocal organs, were denied to exist by the self which made the planchette go. (p. 208)

James discussed hypnosis specifically, and the evidence that individuals who are given directions when in a hypnotic trance to engage in certain actions indeed perform these actions when they are no longer in a trance and have no recollection of the actions having been suggested to them in a trance.

> *We get exactly similar results in the so-called post-hypnotic suggestion*. It is a familiar fact that certain subjects, when told during a [hypnotic] trance to perform an act or to experience an hallucination after waking, will when the time comes, obey the command. How is the command registered? How is its performance so accurately timed? These problems were long a mystery, for the primary personality remembers nothing of the trance or the suggestion, and will often trump up an improvised pretext for yielding to the unaccountable impulse which possesses the man so suddenly and which he cannot resist. Edmund Gurney was the first to discover, by means of automatic writing, that the secondary self is awake, keeping its

> attention constantly fixed on the command and watching for the signal of its execution. (pp. 208-209)

James then combined post-hypnotic trance with "automatic writers," those apparently suffering from hysteria.

> Certain trance-subjects who were also automatic writers, when roused from trance and put to the planchette,–not knowing then what they wrote, and having their upper attention fully engrossed by reading aloud, talking, or solving problems in mental arithmetic,–would inscribe the orders which they had received, together with notes relative to the time elapsed and the time yet to run before the execution. *It is therefore to no 'automatism' in the mechanical sense that such acts are due: a self presides over them, a split-off, limited and buried, but yet a fully conscious, self* [italics added]. More than this, the buried self often comes to the surface and drives out the other self whilst the acts are performing. In other words, the subject lapses into trance again when the moment arrives for execution, and has no subsequent recollection of the act which he has done. Gurney and Beaunis established this fact, which has since been verified on a large scale; and Gurney also showed that the patient became *suggestible* again during the brief time of the performance. M. Janet's observations, in their turn well illustrate the phenomenon. (p. 209)

We see then that James noted that there were two executive agencies, often unaware of each other and each corresponding to Freud's concept of ego, governing their respective psychological agencies. James used Janet's subject Lucie to support his point that post-hypnotic trance behavior demonstrates these two executive agencies. He quoted Janet:

> "I tell Lucie to keep her arms raised after she shall have awakened. Hardly is she in the normal state, when up go her arms above her head, but she pays no attention to them. She goes, comes, converses, holding her arms high in the air. If asked what her arms are doing, she is surprised at such a question, and says very sincerely: 'My hands are doing nothing; they are just like yours.'... I command her to weep, and when awake she really sobs, but continues in the midst of her tears to talk of very gay matters. The sobbing over, there remained no trace of this grief, which seemed to have been quite sub-conscious." (pp. 209-210)

In the following, James expressed his own sense of the unusual character of the behavior of Léonie and Lucie.

> The primary self often has to invent an hallucination by which to mask and hide from its own view the deeds which the other self is enacting. Léonie 3 (M. Janet designates by numbers the different personalities which the subject may display.) writes real letters, whilst Léonie 1 believes that she is knitting; or Lucie 3 really comes to the doctor's office, whilst Lucie 1 believes herself to be at home. This is a sort of delirium. The alphabet, or the series of numbers, when handed over to the attention of the secondary personage may for the time be lost to the normal self. Whilst the hand writes the alphabet, obediently to command, the 'subject,' to her great stupefaction, finds herself unable to recall it, etc. *Few things are more curious than these relations of mutual exclusion, of which all gradations exist between the several partial consciousnesses* [italics added].[7] (p. 210)

James then began to discuss opinions regarding whether these mutually exclusive consciousnesses characterize those of use who are "normal" as well as those who suffer from hysteria. The "normal" mind to this day is generally considered to be unitary in nature integrating disparate experiences within itself and with a cohesive sense that is called one's identity.

> How far this splitting up of the mind into separate consciousnesses may exist in each one of us is a problem. M. Janet holds that it is only possible where there is abnormal weakness, and consequently a defect of unifying or co-ordinating power. An hysterical woman abandons part of her consciousness because she is too weak nervously to hold it together. The abandoned part meanwhile may solidify into a secondary or sub-conscious self. In a perfectly sound subject, on the other hand, what is dropped out of mind at one moment keeps coming back at the next. The whole fund of experiences

---

[7] The text in parentheses appears as a footnote in *The Principles of Psychology*.

> and knowledges remains integrated, and no split-off portions of it can get organized stably enough to form subordinate selves. (p. 210)

Attempting to provide further evidence for the existence of a second consciousness, or executive agency, James provided evidence for certain characteristics of the executive agency that dwelled mostly in the background.

> The stability, monotony, and stupidity of these latter is often very striking. The post-hypnotic sub-consciousness seems to think of nothing but the order which it last received; the cataleptic sub-consciousness, of nothing but the last position imprinted on the limb. M. Janet could cause definitely circumscribed reddening and tumefaction of the skin on two of his subjects, by suggesting to them in hypnotism the hallucination of a mustard-poultice of any special shape. "J'ai tout le temps pensé à votre sinapisme," says the subject, when put back into trance after the suggestion has taken effect. A man N.,...whom M. Janet operated on at long intervals, was betweenwhiles tampered with by another operator, and when put to sleep again by M. Janet, said he was 'too far away to receive orders, being in Algiers.' The other operator, having suggested that hallucination, had forgotten to remove it before waking the subject from his trance, and the poor passive trance-personality had stuck for weeks in the stagnant dream. Léonie's sub-conscious performances having been illustrated to a caller, by a '*pied de nez*' executed with her left hand in the course of conversation, when, a year later, she meets him again, up goes the same hand to her nose again, without Léonie's normal self suspecting the fact. (pp. 210-211)

James, though, appeared to differ from Janet and maintain that these mutually exclusive consciousnesses may characterize anyone.

> *All these facts, taken together, form unquestionably the beginning of an inquiry which is destined to throw a new light into the very abysses of our nature. It is for that reason that I have cited at such length in this early chapter of the book. They prove one thing conclusively, namely, that we must never take a person's testimony, however sincere, that he has felt nothing, as*

> *proof positive that no feeling has been there. It may have been there as part of the consciousness of a 'secondary personage,' of whose experiences the primary one whom we are consulting can naturally give no account.*[8] (p. 211)

Next James focused on the relationship between these distinct consciousnesses. In particular, he pointed out that there is a recognition of at least one consciousness of the other whereby the former consciousness actively excludes some feature of the world from its own experience.

> In hypnotic subjects (as we shall see in a later chapter) just as it is the easiest thing in the world to paralyze a movement or member by simple suggestion, so it is easy to produce what is called a systematized anæsthesia by word of command. A systematized anæsthesia means an insensibility, not to any one element of things, but to some one concrete thing or class of things. The subject is made blind or deaf to a certain person in the room and to no one else, and thereupon denies that that person is present, or has spoken, etc. M. P. Janet's Lucie, blind to some of the numbered cards in her lap (p. 207 above), is a case in point. Now when the object is simple, like a red wafer or a black cross, the subject, although he denies that he sees it when he looks straight at it, nevertheless gets a 'negative after-image' of it when he looks away again, showing that the *optical impression* of it has been received. Moreover reflection shows that such a subject must *distinguish the object from others like it in order to be blind to it*. Make him blind to one person in the room, set all the persons in a row, and tell him to count them. He will count all but that one. But how can he tell which one not to count without recognizing who he is? In like manner, make a stroke on paper or blackboard, and tell him it is not there, and he will see nothing but the clean paper or board. Next (he not looking) surround the original stroke with other strokes exactly like it, and ask him what he sees. He will point out one by one all the new strokes, and omit the original one every time, no matter how numerous the new strokes may be, or

---

[8] *"We must never take a person's testimony, however sincere, that he has felt nothing, as proof positive that no feeling has been there."* is from original text.

> in what order they are arranged. Similarly, if the original single stroke to which he is blind be *doubled* by a prism of some sixteen degrees placed before one of his eyes (both being kept open), he will say that he now sees *one* stroke, and point in the direction in which the image seen through the prism lies, ignoring still the original stroke. (pp. 211-212)

Having discussed the point implied in Freud's writings on splitting of the ego, namely that one consciousness or ego must distinguish the other/s in order to be "blind" to it, James then discussed this very important point in more detail.

> Obviously, then, he is not blind to the *kind* of stroke in the least. He is blind only to one individual stroke of that kind in a particular position on the board or paper–that is to a particular complex object; and, paradoxical as it may seem to say so, he must distinguish it with great accuracy from others like it, in order to remain blind to it when the others are brought near. He discriminates it, as a preliminary to not seeing it at all.
>
> Again, when by a prism before one eye [and only that eye] a previously invisible line [presumably through hypnosis] has been made visible to that eye, and the other eye is thereupon closed or screened, *its* closure makes no difference; the line still remains visible. But if then the prism be removed, the line will disappear even to the eye which a moment ago saw it, and both eyes will revert to their original blind state.
>
> We have, then, to deal in these cases neither with a blindness of the eye itself, nor with a mere failure to notice, but with something much more complex; namely, an active counting out and positive exclusion of certain objects. It is as when one 'cuts' an acquaintance, 'ignores' a claim, or 'refuses to be influenced' by a consideration. But the perceptive activity which works to this result is *disconnected* [italics added] from the consciousness which is personal, so to speak, to the subject, and makes of the object concerning which the suggestion is made, its own private possession and prey. (pp. 212-213)

Notice that the consciousness that employs this "blindness" to some particular feature of the world is not aware of its own blindness. These consciousnesses are mutually exclusive since one is concerned with the

On Whether People

occurrence of some experience and the other is concerned with its denial.  It is really to say that two executive agencies are functioning in one mind.  As if to reinforce the radical nature of the these he discussed, James wrote in a footnote:

> How to conceive of this state of mind is not easy. It would be much simpler to understand the process, if adding new strokes made the first one visible. There would then be two different objects apperceived as totals,–paper with one stroke, paper with many strokes; and, blind to the former, he would see all that was in the latter, because he would have apperceived it as a different total in the first instance.
>
> A process of this sort occurs sometimes (not always) when the new strokes, instead of being mere repetitions of the original one, are lines which combine with it into a total object, say a human face. The subject of the trance then may regain his sight of the line to which he had previously been blind, by seeing it as part of the face. (p. 213)

James was struggling in the above quote with the mutual exclusivity of the two different consciousnesses, implying two different executive functions. He tried to show that if some perceptual totality underlaid the phenomena he discussed, one could argue that there is but one executive mental agency and that the apparently mutually exclusive phenomena reflect are perceptual wholes of which they are part.  He discussed the human face in this regard.  It should be remembered that James maintained, difficult as it was for him, that distinct and different consciousnesess could simultaneously exist for the same person.

James concludes by giving an example from everyday life for a normal person of the phenomenon discussed.

> The mother who is asleep to every sound but the stirrings of her babe, evidently has the babe-portion of her auditory sensibility systematically awake. Relatively to that, the rest of her mind is in a state of systematized anæsthesia. That department, split  off and disconnected from the sleeping part, can none the less wake the latter up in case of need. So that on the whole the quarrel between Descartes and Locke as to whether the mind ever sleeps is less near to solution than ever. On *a priori* speculative grounds Locke's view that thought and feeling may at times wholly disappear seems the more plausible.

> As glands cease to secrete and muscles to contract, so the brain should sometimes cease to carry currents, and with this minimum of its activity might well coexist a minimum of consciousness. *On the other hand, we see how deceptive are appearances, and are forced to admit that a part of consciousness may sever its connections with other parts and yet continue to be* [italics added]. On the whole it is best to abstain from a conclusion. The science of the near future will doubtless answer this question more wisely than we can now. (p. 213)

Whether James's questions concerning whether the mind sleeps has or has not been answered conclusive is secondary to the point that evidence comes from physics that mutually exclusive consciousnesses exist, supporting the findings of James and Freud.

### THE CONTEMPORARY CONSIDERATION OF HYSTERIA

In order to demonstrate that the mental disorder that James referred to as hysteria is present today, some quotes are presented from the *Diagnostic and Statistical Manual of Mental Disorders* (1994), known as *DSM-IV*. Today, hysterical symptoms are considered within two disorders: conversion disorder and dissociative identity disorder.

*Conversion Disorder*

> Following are quotes from the *DSM-IV* on conversion disorder.

> Conversion Disorder involves unexplained symptoms or deficits affecting voluntary motor or sensory function that suggest a neurological or other general medical condition. Psychological factors are judged to be associated with the symptoms or deficits....

> Conversion symptoms typically do not conform to known anatomical pathways and physiological mechanisms, but instead follow the individual's conceptualization of a condition. A "paralysis" may involve inability to perform a particular movement or to move an entire body part, rather than a deficit corresponding to patterns of motor innervation. Conversion symptoms are often inconsistent. A "paralyzed" extremity will be moved inadvertently while dressing or when attention is directed elsewhere. If placed above the head and released, a "paralyzed" arm will briefly retain its position, then fall to the side, rather than striking the head. Unacknowledged strength in antagonistic muscles, normal muscle tone, and intact reflexes may be demonstrated. An electromyogram will be normal. Difficulty swallowing will he equal with liquids and solids. Conversion "anesthesia" of a foot or a hand may follow a so-called stocking-glove distribution with uniform (no proximal to distal gradient) loss of all sensory modalities (i.e., touch, temperature, and pain) sharply demarcated at an anatomical landmark rather than according to dermatomes. A conversion "seizure" will vary from convulsion to convulsion, and paroxysmal activity will not be evident on an EEG....

Conversion symptoms are related to voluntary motor or sensory functioning and are thus referred to as "pseudoneurological." Motor symptoms or deficits include impaired coordination or balance, paralysis or localized weakness, aphonia, difficulty swallowing or a sensation of a lump in the throat, and urinary retention. Sensory symptoms or deficits include loss of touch or pain sensation, double vision, blindness, deafness, and hallucinations. Symptoms may also include seizures or convulsions. The more medically naive the person, the more implausible are the presenting symptoms. More sophisticated persons tend to have more subtle symptoms and deficits that may closely simulate neurological or other general medical conditions....

Reported rates of Conversion Disorder have varied widely, ranging from 11/100,000 to 300/100,000 in general population samples. It has been reported as a focus of treatment in 1%-3% of outpatient referrals to mental health clinics. (*Diagnostic and Statistical Manual of Mental Disorders*, 1994, pp. 445, 452-455)

*Dissociative Identity Disorder*

Following are quotes from the DSM-IV on dissociative identity disorder.

Dissociative Identity Disorder (formerly Multiple Personality Disorder) is characterized by the presence of two or more distinct identities or personality states that recurrently take control of the individual's behavior accompanied by an inability to recall important personal information that is too extensive to be explained by ordinary forgetfulness....

Dissociative Identity Disorder reflects a failure to integrate various aspects of identity, memory, and consciousness. Each personality state may he experienced as if it has a distinct personal history, self-image, and identity, including a separate name. Usually there is a primary identity that carries the individual's given name and is passive, dependent, guilty, and depressed. The alternate identities frequently have different names and characteristics that contrast with the primary identity (e.g., are hostile, controlling, and self-destructive). Particular identities may emerge in specific circumstances and may differ in reported age and gender, vocabulary, general knowledge, or predominant affect. Alternate identities are experienced as taking control in sequence, one at the expense of the other, and may deny knowledge of one another, be critical of one another, or appear to be in open conflict. Occasionally, one or more powerful identities allocate time to the others. Aggressive or hostile identities may at times interrupt activities or place the others in uncomfortable situations.

Individuals with this disorder experience frequent gaps in memory for personal history, both remote and recent. The amnesia is frequently asymmetrical. The more passive identities tend to have more constricted memories, whereas the more hostile, controlling, or 'protector" identities have more complete memories. An identity that is not in control may nonetheless gain access to consciousness by producing auditory or visual hallucinations (e.g., a voice giving instructions). Evidence of amnesia may be uncovered by reports from others who have witnessed behavior that is disavowed by the individual or by the individual's own discoveries (e.g., finding items of clothing at home that the individual cannot remember having bought). There may be loss of memory not only for recurrent periods of time, but also an overall loss of biographical memory for some extended period of childhood. Transitions among identities are

> often triggered by psychosocial stress. The time required to switch from one identity to another is usually a matter of seconds, but, less frequently, may be gradual. The number of identities reported ranges from 2 to more than 100. Half of reported cases include individuals with 10 or fewer identities....
>
> The sharp rise in reported cases of Dissociative Identity Disorder in the United States in recent years has been subject to very different interpretations. Some believe that the greater awareness of the diagnosis among mental health professionals has resulted in the identification of cases that were previously undiagnosed. In contrast, others believe that the syndrome has been overdiagnosed in individuals who are highly suggestible. (*Diagnostic and Statistical Manual of Mental Disorders*, 1994, pp. 477, 484-486)

The problem discussed by Einstein, Podolsky, and Rosen on the possibility of there existing simultaneously mutually exclusive situations in the physical world has been noted. Now the root of this problem in terms of the nature of the wave function in quantum mechanics will be discussed.

### How the Nature of the Wave Function in Quantum Mechanics Underlies the Problem Posed by Einstein, Podolsky, and Rosen

How the nature of the wave function is central to the problem posed by Einstein, Podolsky, and Rosen in 1935 will be discussed through a presentation of Einstein's view of the wave function in quantum mechanics. Einstein (1949/1969) addressed the relevant broad principles of quantum mechanics as well as the argument he developed with Podolsky and Rosen in his "Autobiographical Notes." Bohr's response to this problem is discussed as it is central to understanding the underlying issues.

Einstein first noted that Newtonian mechanics is readily understood in terms of the realistic basis of physics.

> Physics is an attempt conceptually to grasp reality as it is thought independently of its being observed. In this sense one speaks of "physical reality." In pre-quantum physics there was no doubt as to how this was to be understood. In Newton's theory reality was determined by a material point in space and

## On Whether People

time [functioning in a deterministic manner independent of cognition]; in Maxwell's theory, by the field in space and time. (pp. 81, 83)

Einstein (1949/1969) continued:

> In quantum mechanics it is not so easily seen [i.e., the realistic basis of physics]. If one asks: Does a $\Psi$-function of the quantum theory represent a real factual situation in the same sense in which this is the case of a material system of points or of an electromagnetic field, one hesitates to reply with a simple "yes" or "no"; why? What the $\Psi$-function (at a definite time) asserts, is this: What is the probability for finding a definite physical magnitude $q$ (or $p$) [of a physical system] in a definitely given interval, if I measure it at time $t$? The probability is here to be viewed as an empirically determinable, and therefore certainly as a "real" quantity which I may determine if I create the same $\Psi$-function very often and perform a $q$-measurement each time. But what about the single measured value of $q$? Did the respective individual system have this $q$-value even before the measurement? To this question there is no definite answer within the framework of the [existing] theory, since the measurement is a process which implies a finite disturbance of the system from the outside [generally resulting in the change in wave function that occurs immediately throughout space]; it would therefore be thinkable that the system obtains a definite numerical value for $q$ (or $p$), i.e., the measured numerical value, only through the measurement itself. (p. 83)[9]

Then Einstein presented the essence of a gedankenexperiment that he had proposed earlier with Podolsky and Rosen (Einstein, Podolsky, and Rosen, 1935).

> We now present...the following instance: There is to be a system which at the time $t$ of our observation consists of two partial systems $S_1$ and $S_2$, which at this time are spatially separated [without limit on the separation] and (in the sense of the classical physics) are without significant reciprocity. The total system is to be completely described through a known $\Psi$-function $\Psi_{12}$ in the sense of quantum mechanics. All quantum theoreticians now agree upon the following: If I make a complete measurement of $S_1$, I get from the results of the measurement and from $\Psi_{12}$ an entirely definite $\Psi$-function $\Psi_2$ of the system $S_2$ [immediately]. The character of $\Psi_2$ then depends upon *what kind* of measurement I undertake on $S_1$.

---

[9] The term *existing*, along with the brackets that enclose it, that are found in the quote are actually part of the quoted material and not added by myself.

Now it appears to me that one may speak of the real factual situation of the partial system $S_2$. Of this real factual situation, we know to begin with, before the measurement of $S_1$, even less than we know of a system described by the $\Psi$-function. But on one supposition we should, in my opinion, absolutely hold fast: the real factual situation of the system $S_2$ is independent of what is done with the system $S_1$, which is spatially separated from the former. According to the type of measurement which I make of $S_1$, I get, however, a very different $\Psi_2$ for the second partial system ($\Psi_2$, $\Psi_2^1$,...). Now, however, the real situation of $S_2$ must be independent of what happens to $S_1$. For the same real situation of $S_2$ it is possible therefore to find, according to one's choice, different types of $\Psi$-function. (One can escape from this conclusion only by either assuming that the measurement of $S_1$ ((telepathically)) changes the real situation of $S_2$ or by denying independent real situations as such to things which are spatially separated from each other. Both alternatives appear to me entirely unacceptable.)

If now...physicists...accept this consideration as valid, then *B* [a particular physicist] will have to give up his position that the $\Psi$-function constitutes a complete description of a real factual situation. For in this case [i.e., the case of a complete description] it would be impossible that two different types of $\Psi$-functions [representing mutually exclusive situations] could be co-ordinated [simultaneously] with the identical factual situation of $S_2$ [the same concrete physical circumstances]. (Einstein, 1949/1969, pp. 85; 87)

# On Whether People

Bohr's (1935) response to Einstein, Podolsky, and Rosen's gedankenexperiment was that there is an unavoidable interaction between the physical existent measured and the measuring instrument in their gedankenexperiment that cannot be ignored. Essentially, Bohr's response was that the situation Einstein, Podolsky, and Rosen were referring to is quantum mechanical in its structure. That is, the structure of the gedankenexperiment presented by Einstein, Podolsky, and Rosen was based on: (1) probabilistic prediction rooted in the quantum mechanical wave function that describes the physical system, and (2) the in general immediate change throughout space of the wave function upon measurement of the physical system.

Furthermore, Bohr was saying that because the situation described by Einstein, Podolsky, and Rosen is framed within the theory of quantum mechanics, their result that two very different wave functions (really two mutually exclusive views of the world) can characterize the same concrete physical circumstances in quantum mechanics is incorrect. According to Bohr, the particular interaction of the measuring apparatus and $S_1$ is associated with a specific state of $S_2$ upon the measurement of $S_1$. But Einstein, Podolsky, and Rosen's result is basically correct. Two very different wave functions can indeed characterize the same concrete physical circumstances, even if the state of $S_2$ depends on the measurement result at $S_1$. The velocity limitation of the special theory of relativity, the velocity of light in vacuum, is not a limiting factor in the change of the wave function throughout space when a measurement is made.

Where Bohr was correct was in noting that the conception of physical reality that they indeed adopted in their gedankenexperiment was that "physics is an attempt conceptually to grasp reality as it is thought independently of its being observed" (Einstein, 1949/1969, p. 81), and this conception of physical reality is not part of quantum mechanics. Yet it is quantum mechanics that Einstein, Podolsky, and Rosen used to structure their gedankenexperiment. According to Bohr, in not allowing for the interaction between the physical existent measured and the measuring process, of which the observer is the chief component, in defining an element of physical reality, they were able to frame their argument so as to obtain the result that quantum mechanics is not a complete theory of the physical world. Thus, Bohr was correct in his criticism up to a point, and Einstein, Podolsky, and Rosen were correct without the artificial constraint of their realistic definition of the physical world and its essential independence of the physical theory describing it.

# On Whether People

<small>JAMES'S INFLUENCE ON BOHR</small>

In the mid-1920's, Bohr came to believe that in quantum mechanics, certain quantities (such as the position and momentum) of certain physical existents (such as an electron) cannot both be known with arbitrary precision. He argued that in principle descriptions of these quantities are mutually exclusive. In that the mutually exclusive descriptions of these quantities could both describe the existent and as these descriptions together could simultaneously apply to the existent in pre-quantum physics, Bohr called these descriptions complementary. Bohr anchored complementarity to the physical world because, for Bohr, the mutually exclusive descriptions were determined by the concrete experimental arrangements that the physicist had selected (e.g., one experimental arrangement to measure position and another experimental arrangement to measure momentum of an electron) (Bohr, 1935).

Bohr was significantly influenced by William James in his development of the concept of complementarity, a central concept in the theory of quantum mechanics. Jammer (1966/1989) argued that James's work had a significant impact on Bohr's work in physics, specifically in his development of complementarity. Jammer noted, "Bohr repeatedly admitted how impressed he was particularly by the psychological writings of this American philosopher" (p. 182). It appears that Bohr was well-acquainted with certain ideas discussed in James's *The Principles of Psychology*. Jammer argued that it was probably James's use of the term complementarity in the context of his discussion of work on hysteria that we have reviewed that had the major impact on Bohr.

In contrast to Bohr's above stated view of complementarity, Bohr (1934/1961) seems to have suggested that complementarity itself might fundamentally involve the fundamental structuring of perception, namely the essential separation between that which is perceived and the perceiving person. In this separation only a part of the world is accessible to the perceiving person because this person of necessity maintains a particular stance in the world.

> For describing our mental activity [which includes perceptions of the physical world], we require, on one hand, an objectively given content to be placed in opposition to a perceiving subject, while, on the other hand, as is already implied in such an assertion, no sharp sensation between object and subject can be maintained, since the perceiving subject also belongs to our mental content. From these circumstances follows...that a

> complete elucidation of one and the same object may require diverse points of view which defy a unique description. (Bohr, 1934/1961, p. 96)

Consider the following quote. Here Bohr appeared to attribute the extent of the realm of the measurable space upon which physics depends, and rooted in basic psychological experience, to whether a physical entity is considered part of the physical world that can be measured by an observer or instead is an extension of the human observer who is attempting to measure the physical world.

> It is very instructive that already in simple psychological experiences we come upon fundamental features not only of the relativistic but also of the reciprocal [complementary] view. The relativity of our perception of motion, with which we become conversant as children when travelling by ship or by train, corresponds to common-place experiences on the reciprocal character of the perception of touch. One need only remember here the sensation, often cited by psychologists, which every one has experienced when attempting to orient himself in a dark room by feeling with a stick. When the stick is held loosely, it appears to the sense of touch to be an object. When, however, it is held firmly, we lose the sensation that it is a foreign body, and the impression of touch becomes immediately localized at the point where the stick is touching the body under investigation. It would scarcely be an exaggeration to maintain, purely from psychological experiences, that the concepts of space and time by their very nature acquire a meaning only because of the possibility of neglecting the interaction with the means of measurement. (Bohr, 1934/1961, pp. 98-99)

It appears that the inescapable interaction between the measuring apparatus and the physical entity measured that was at the heart of Bohr's concept of complementarity might indeed be subsumed in the more fundamental structure of perception.

In addition, with a bit more attention to James's description of hysteria, Bohr may well have recognized that Einstein, Podolsky, and Rosen's experiment really afforded the possibility of mutually exclusive situations characterizing the same concrete physical circumstances. As discussed, James (1899) acknowledged that the hysteric manifests "possible consciousnesses...[that nonetheless may] coexist....[even though you may] give an object to one of the consciousnesses, and by that fact you remove it from the other or others" (p. 204).

## Conclusion

Einstein, Podolsky, and Rosen showed that in quantum mechanics two different wave functions predicting specific values for quantities represented by non-commuting Hermitian operators can characterize a physical existent. This raises the question whether observations of mutually exclusive physical phenomena are possible. Evidence from psychology has been presented that indicates that people may indeed have the capacity to make such observations. There is additional evidence supporting this conclusion from research on adaptation to inversion of incoming light.

The evidence from psychology presented here in large part stemmed originally from the study of mental illness. Both Freud and James saw the relevance of their insights to normal mental functioning as well. One might question the usefulness of insights originally gained from a study of mental phenomena characterizing a small percentage of people. But this circumstance is not different than situations concerning physical phenomena where broad physical principles are developed initially on the basis of evidence from physical phenomena that at least at first are not frequently encountered and seem to have little to do with understanding the vast majority of physical phenomena that we encounter in our daily lives. Thus, stellar aberration, the propagation of light in moving media, the invariant velocity of light in vacuum, and even what appeared as the curious properties of electric and magnetic fields in Einstein's day were significant factors in the development and verification of the special theory of relativity (Resnick, 1968). The odd findings that subatomic particles had wave-like properties, and that light had particle-like properties led to quantum theory, a theory much more powerful than classical physics, including Newtonian mechanics, in understanding physical phenomena.

# On Whether People

On Whether People

Princeton, New Jersey: Princeton University Press. (Original work published 1961)